# Shallow Depth Factoring Based on Quantum Feasibility Labeling and Variational Quantum Search

Imran Khan Tutul[1], Sara Karimi[2], Mohammadreza Soltaninia[1], Junpeng Zhan[1*]

[1]Department of Electrical Engineering, Alfred University, Alfred, NY 14802, USA

[2]Department of Mechanical Engineering, Alfred University, Alfred, NY 14802, USA

**Abstract**

Large integer factorization is a prominent research challenge, particularly in the context of quantum computing. This holds significant importance, especially in information security that relies on public key cryptosystems. The classical computation of prime factors for an integer has exponential time complexity. Quantum computing offers the potential for significantly faster computational processes compared to classical processors. In this paper, we propose a new quantum algorithm, Shallow Depth Factoring (SDF), to factor a biprime integer. SDF consists of three steps. First, it converts a factoring problem to an optimization problem without an objective function. Then, it uses a Quantum Feasibility Labeling (QFL) method to label every possible solution according to whether it is feasible or infeasible for the optimization problem. Finally, it employs the Variational Quantum Search (VQS) to find all feasible solutions. The SDF utilizes shallow-depth quantum circuits for efficient factorization, with the circuit depth scaling linearly as the integer to be factorized increases. Through minimizing the number of gates in the circuit, the algorithm enhances feasibility and reduces vulnerability to errors.

## I. Introduction

Quantum computing has emerged as an exciting new field that seeks to harness the principles of quantum mechanics to perform computational tasks that are beyond the capabilities of classical computers. Recent advances in quantum hardware have led to the construction of quantum computers with large numbers of qubits. For example, in 2019 IBM created the 27-qubit Falcon quantum computer and in the next year 65-qubit Hummingbird was introduced. The 433-qubit Osprey has the largest number of qubit and was introduced by IBM in 2022 and the company plans to build a 1000-qubit quantum computer by 2024. In addition to the fast development of the quantum hardware, quantum algorithms, for example, pure quantum algorithms such as Shor's algorithm (Shor 1994) and Grover's algorithm (Grover 1996), quantum simulation algorithms that can be used to simulate quantum systems (Ortiz et al. 2001), variational quantum algorithms that use a hybrid classical-quantum approach (Peruzzo et al. 2014; McArdle et al. 2018), quantum error correction algorithms (Cai et al. 2021; Reed et al. 2012), have already demonstrated significant advantages in areas such as cryptography, optimization, and simulation, with potential applications in fields ranging from chemistry (Cao et al. 2019) and materials science (Yost et al. 2020) to finance (Stamatopoulos et al. 2020) and machine learning (Biamonte et al. 2016).

Prime factorization is a fundamental problem in computer science and cryptography that involves decomposing a composite number into its prime factors. While this task can be accomplished efficiently using classical algorithms for small numbers, the time required to factor large integers increases exponentially with the number of digits. The security of cryptocurrencies, such as Bitcoin and Ethereum, depends on cryptographic hash functions that employ prime numbers. Therefore, factoring large numbers could potentially compromise the security of the entire cryptocurrency network. The development of an efficient algorithm for factoring large numbers would pose a significant threat to information security in the digital age, as it would render encryption schemes vulnerable to attacks. This would have profound effects on the protection of sensitive data such as financial transactions, personal information, and government secrets.

The problem of prime factorization has been of interest to mathematicians for centuries. The earliest known algorithm for factoring integers was developed by Euclid over 2000 years ago (Euclid 2002; Pettofrezzo and Byrkit 1970), and it remains in use today due to its simplicity and efficiency. Euclid's algorithm is a sequential process that repeatedly divides the input number by smaller primes until the remaining factor is prime. However, due to its exponential time complexity, Euclid's algorithm is limited in its ability to factor large numbers. In the 20th century, several more advanced algorithms were developed for prime factorization, including the Pollard-Rho algorithm (Pollard 1975), the Quadratic Sieve algorithm (Pomerance 1982), and the General Number Field Sieve algorithm (Lenstra and Hendrik Jr 1993). The Pollard-Rho algorithm (Pollard 1975), proposed in 1975, is a probabilistic algorithm that uses random numbers to search for a non-trivial factor of a composite number. While it is more efficient than Euclid's algorithm for large numbers, it is still not practical for factoring numbers with hundreds or thousands of digits. The

*Corresponding Author: J.Z. E-mail: zhanj@alfred.edu; zhanjunpeng@gmail.com

Quadratic Sieve algorithm (Pomerance 1982), introduced in 1981, is a deterministic algorithm that uses modular arithmetic and linear algebra to factor large integers. This is more efficient than the Pollard-Rho algorithm (Pollard 1975), however, its time complexity is still too high for factoring very large numbers. Finally, the General Number Field Sieve algorithm (Pettofrezzo and Byrkit 1970; Mollin 2002), developed in the 1990s, is currently the fastest known classical algorithm for factoring large integers. Despite its efficiency, the time required to factor large numbers using this algorithm is still exponential and increases rapidly with the size of the input.

Quantum computing is motivated by limitations of classical computers in solving certain problems, particularly those related to cryptography and number theory. Prime factorization (Aoki et al. 2007; Long 1972; Hardy 2008) is a key problem in number theory. It can be approached from two major perspectives, namely, factoring a number based on Shor's algorithm and factorization as an optimization problem. One of the most notable quantum algorithms is Shor's algorithm, which can efficiently factor integers up to a few hundred digits long. This is one of the quantum algorithms that allow quantum computers to outperform classical supercomputers. However, the numbers that have been factored using Shor's algorithm on actual quantum computers to date have been relatively small, with the largest being only 21 (Skosana and Tame 2021). The primary challenge in implementing Shor's algorithm lies in the modular arithmetic part, which requires a large number of qubits and operations with high accuracy. Furthermore, the algorithm's circuit depth and complexity, along with the need for error correction, make it challenging to implement on current quantum hardware. In 2013 (Gamel and James 2013), a paper based on a compiled version of Shor's algorithm proposed a verification scheme to avoid the bottleneck of the Shor's algorithm, modular exponentiation. However, when the values of the base, exponent, and modulus, which are parameters of the modular exponentiation become more complex or larger, the task of designing an efficient circuit for modular exponentiation becomes much more difficult using existing technologies. This method (Gamel and James 2013; Geller and Zhou 2013) is interesting only when it comes with simple intermediate steps which provide a significant simplification of a complex periodic (modular exponentiation) circuit. The compilation process relies on known information, such as the period or order of the solution, to simplify the circuit from its complex general form to a more manageable size. This limits the performance of the method. Nevertheless, Shor's algorithm remains a potent demonstration of quantum computing's potential and has motivated further research into developing more efficient and scalable quantum algorithms.

The second approach converts the factorization problem into an optimization problem and solves it using methods like Adiabatic quantum computation and quantum annealing principles. The Adiabatic quantum computation method (Farhi et al. 2000; Peng et al. 2008) utilizes optimization in the preprocessing part. This preprocessing part can be subsequently reduced to a set of equations. Complex Hamiltonian is derived from these equations and encodes the solution in its ground state. Number 143 was factored out using this method which required 4 qubits (Xu et al. 2011). In terms of complexity, although the implementation time of this method shows polynomial behavior with the size of the problem, the proof for the complexity of the adiabatic quantum algorithm remains an open question (Peng et al. 2008). In 2018, a group of authors proposed a new method, variational quantum factoring (VQF) (Anschuetz et al. 2019). The VQF simplifies equations over Boolean variables in a preprocessing step to reduce the required qubit count. It then employs variational circuits trained with the quantum approximate optimization algorithm (QAOA) to approximate the ground state of the resulting Ising Hamiltonian. However, the limitation of the mentioned approach is that it assumes prior knowledge of certain information, such as the bit length of the factors. In 2020 (Wang et al. 2020), another method for decoding RSA was proposed by a group of authors who successfully factored all integers within 10000 using the optimistic capability of a D-wave quantum computer. A D-Wave using quantum annealing provides a novel approach that demonstrates significant advantages of factoring 20-bit integers (1028171). Moreover, their method requires more qubits than Shor's algorithm. Its complexity is $\mathcal{O}(log^2(N))$, where $N$ is the number to be factored and, the Shor's algorithm is better than their approach in terms of theorical complexity.

In 2018, a new approach utilizing Grover's algorithm with a combination of classical algorithm/analytic algebra (Dash et al. 2018) was proposed for factoring integers. The authors employed IBM's 5- and 16-qubit quantum processors to find the factors of two integers, 4088459 and 966887. In their work, they employed a pre-processing step using the minimization method, similar to the adiabatic approach, followed by directly implementing a unitary operation. The unitary operation is an exponential function of the non-unitary Hamiltonian, which is also used in the adiabatic method. These numbers were deliberately chosen to demonstrate that the complexity of the factorization problem through this approach is not contingent on the largeness of the number. Instead, it relies on a particular characteristic of the factors involved.

Recently, a novel algorithm (Yan et al. 2022) has been introduced. It can factor integers up to 48 bits using 10 superconducting qubits, marking the largest integer factored on a quantum computer to date. The algorithm combines classical and quantum parts, utilizing the lattice reduction method for the classical part and the quantum approximate optimization algorithm (QAOA) for the quantum part. However, the classical part of the algorithm involves solving the closest vector problem (CVP) on a lattice, which is a well-known NP-hard problem. The number of qubits required for their method is $\mathcal{O}(\log N / (\log \log N))$ (Yan et al. 2022).

This paper presents a novel approach for factorizing biprime integers, which utilizes both the quantum labeling algorithm proposed in (Zhan 2023b) and the variational quantum search (Zhan 2022). The main contributions of this paper are: (1) the introduction of a novel approach, Shallow Depth Factoring (SDF), to factoring biprime integer numbers, which combines three distinct schemes, namely factoring as optimization, QFL, and VQS, (2) the implementation of a method named bit length estimation that operates without requiring prior knowledge of the bit length of the factors associated with the given integer, and (3) design of a shallow depth QFL quantum circuit that accommodates the constraints of the optimization process. The effectiveness of this approach is demonstrated by successfully factoring both integers 143 and 391, with the potential to extend to larger numbers. In Section II, the method and circuit construction procedure are detailed. Results of the experiments are presented in Section III. The conclusion is drawn in Section IV.

## II. Method

This section describes the SDF method, comprising three main steps. The first step is to convert factoring into an optimization problem without objective function, which is described in Section II-A. The second and third steps are QFL and VQS, which are discussed in more detail in Section II-B. Sections II-C and II-D illustrate how to construct quantum circuits using the QFL for two examples, respectively. Section II-E analyzes the complexity of the SDF.

### A. Factoring as Optimization

Optimization model: Integer factoring can be converted to binary optimization, as described in (Burges 2002). We consider a biprime integer number $N$ to find its factors, represented as $N = p \times q$, where $p$ and $q$ represent prime factors of $N$. We can express binary representation of $N$, $p$, and $q$ as:

$$N = \sum_{k=0}^{n_N - 1} 2^k N_k \quad (1)$$

$$p = \sum_{k=0}^{n_p - 1} 2^k p_k \quad (2)$$

$$q = \sum_{k=0}^{n_q - 1} 2^k q_k \quad (3)$$

where $N_k, p_k, q_k \in \{0, 1\}$ represent the $k^{th}$ bit and $n_N, n_p, n_q$ are the binary bit length of $N$, $p$, and $q$, respectively. Table 1 illustrates binary multiplication operation. The first row of Table 1 is the notion of binary arithmetic. We can construct clause $C_i$ using the following equation:

$$C_i = \sum_{j=0}^{i} q_j p_{i-j} + \sum_{j=0}^{i} z_{j,i} - N_i - \sum_{j=1}^{n_c} 2^j z_{i,i+j} \quad (4)$$

where $0 \leq i \leq n_c$ and $z_{i,j} \in \{0, 1\}$ represents the carry bit from bit position $i$ into $j$ (see Appendix A for details). When $i \geq n_p$ or $i \geq n_q$, the terms $q_j p_{i-j}$ in (4) become 0. Based on Table 1, $n = n_p + n_q - 1 + k$, where $k$ is the number of additional columns required for carry bits. A total of $(n_p + n_q - 1)$ equations can be obtained from Table 1.

$$p_1 q_0 + p_0 q_1 = N_1 + 2z_{1,2} \quad (5)$$
$$p_2 q_0 + p_1 q_1 + p_0 q_2 + z_{1,2} = N_2 + 2z_{2,3} + 4z_{2,4} \quad (6)$$
$$\vdots$$
$$z_{(n-1),n} + z_{(n-2),n} = N_{n_N - 1} \quad (7)$$

We can further simplify these equations and construct simplified clauses using pre-processing rules outlined in

Table 1: Binary Multiplication.

| | $2^n$ | ... | ... | ... | ... | ... | ... | $2^2$ | $2^1$ | $2^0$ |
|---|---|---|---|---|---|---|---|---|---|---|
| $p$ | | | | | | $p_{n_p-1}$ | ... | $p_2$ | $p_1$ | $p_0$ |
| $q$ | | | | | | $q_{n_q-1}$ | ... | $q_2$ | $q_1$ | $q_0$ |
| | | | | | | $p_{n_p-1}q_0$ | ... | $p_2 q_0$ | $p_1 q_0$ | $p_0 q_0$ |
| | | | | | $p_{n_p-1}q_1$ | ... | $p_2 q_1$ | $p_1 q_1$ | $p_0 q_1$ | |
| | | | | $p_{n_p-1}q_2$ | ... | $p_2 q_2$ | $p_1 q_2$ | $p_0 q_2$ | | |
| | | | ... | ... | ... | ... | ... | | | |
| | | $p_{n_p-1}q_{n_q-1}$ | ... | $p_2 q_{n_q-1}$ | $p_1 q_{n_q-1}$ | $p_0 q_{n_q-1}$ | | | | |
| Carry | $z_{(n-1),n}$ | ... | ... | ... | ... | $z_{3,4}$ | $z_{2,3}$ | $z_{1,2}$ | | |
| | $z_{(n-2),n}$ | ... | ... | ... | ... | $z_{2,4}$ | | | | |
| | $N_{n_N-1}$ | $N_{n_N-2}$ | ... | ... | ... | ... | ... | | $N_1$ | $N_0$ |

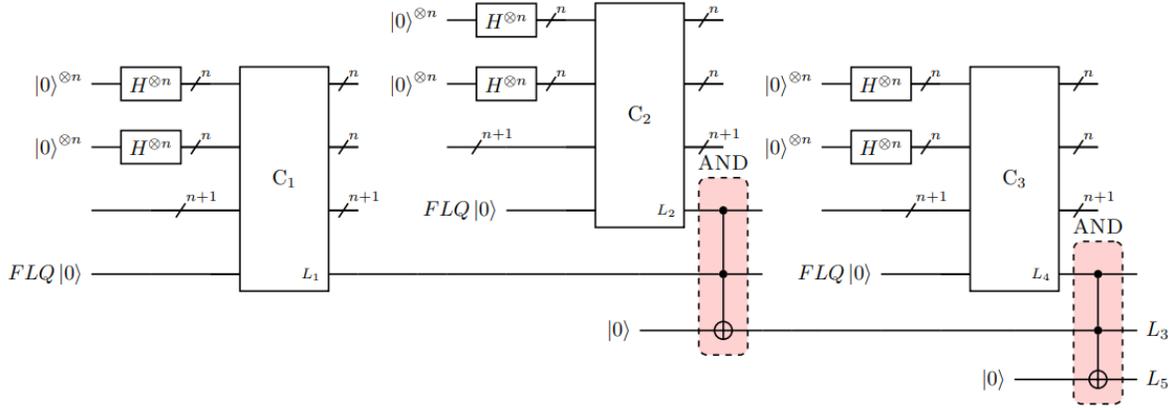

Figure 1: Quantum Feasibility Labeling Circuit for a General Factoring Problem

Appendix A to reduce the complexity of calculations by eliminating some variables and equations.

**Bit length estimation**: The proposed method offers a distinct advantage in that it does not require prior knowledge of the bit length of p and q, which sets it apart from certain previous works (Xu et al. 2011; Dash et al. 2018). Instead, our method uses bit length estimation described in this section to reduce the computational complexity. As $N=p\times q$ and $N$ is biprime, it follows that one of the factors, $n_p$ and $n_q$, must be at least 2 while the other can be at most $(n-1)$-bit number to satisfy $N=p\times q$.

We also find that at least one of the factors ($p$ and $q$) has to be equal to or less than $\sqrt{N}$, which can be expressed as $p\leq \sqrt{N}$ or $q\leq \sqrt{N}$. We can prove this via proof by contradiction. Assume $p>\sqrt{N}$ and $q>\sqrt{N}$, then we have $p\times q > \sqrt{N}\times \sqrt{N}$, that is $p\times q > N$, which contradicts $N=p\times q$. Therefore, we know that either $p$ or $q$ will be equal to or less than $\sqrt{N}$.

Now we show an illustrative example to enhance the understanding of bit length estimation. We use an 8-bit number $N=143$. As per bit length estimation, one of $p$ and $q$ is less or equal to $\sqrt{143}=11.96$ (approximately equal to a 4-bit integer). Since $p$ or $q$ must be less than or equal to $\sqrt{143}$ (a 4-bit number) and 143 cannot be divided by any number with 2 and 3 bits, we can determine that $n_p$ is 4. Now consider choosing $n_q$, it can range anywhere from 2 to 7. When multiplying a 4-bit binary number by a 2-bit binary number, the maximum value that can be obtained is 45. This can be achieved when multiplying 11 (the largest 2-bit number) by 1111 (the largest 4-bit number), resulting in 101101 (45 in decimal). Similarly, when multiplying a 4-bit binary number by a 3-bit binary number, the maximum value that can be obtained is 105. This can be achieved when multiplying 111 (the largest 3-bit number) by 1111 (the largest 4-bit number), resulting in 1101001 (105 in decimal). Since 45 and 105 are less than 143, $q$ cannot be 2 and 3 in bit length. When $q$ is considered as a 4-bit number, the minimum and maximum possible values are 64 and 225, respectively (1000×1000 = 1000000 and 1111×1111= 11100001). As we observe that 143 falls within the range of 64-225, without loss of generality we can consider both $n_p$ and $n_q$ are equal to 4.

**B. Quantum Feasibility Labeling and Variational Quantum Search in Shallow Depth Factoring**

Before delving into the SDF, we first briefly describe the VQS and QFL here. The VQS, introduced in (Zhan 2022), has the remarkable ability to identify good element(s) from an unstructured database, achieving exponential speedup in terms of circuit depth for up to 26 qubits. VQS with a single layer of Ry gates as its Ansatz exhibits near-perfect reachability, making it a promising candidate for efficiently solving NP-complete problems and potentially providing an exponential advantage, in terms of circuit depth, over Grover's algorithm for any number of qubits beyond 26 (Zhan 2023a). While its performance beyond 26 qubits is promising, the trainability of the VQS needs further validation.

In (Zhan 2023b), the author proposed a QFL algorithm that efficiently assigns labels to all potential solutions for the vertex coloring problem. QFL generates a feasible label, expressed as state $|0\rangle$ or $|1\rangle$ of a single qubit, indicating whether the solution is infeasible or feasible for all constraints, respectively. With the labels generated by QFL and the associated possible solutions as input, VQS can effectively identify all feasible solutions of vertex coloring problem.

In the SDF, we utilize both QFL and VQS to identify feasible solutions for simplified clauses derived from binary optimization of the factoring problem, as detailed in the rest of this subsection. Note that the QFL in this paper shares a similar structure with the QFL in (Zhan 2023b), but the circuits are different.

**QFL**: Assuming that after simplifying, $m$ clauses are required to be satisfied for obtaining the prime factor. We employ a C module (represented as $C_i$, $1\leq i\leq m$, as shown in Figure 1 where *m*=3) to represent a simplified clause of the factoring problem. Each C module uses a single qubit (denoted as $L_i$ in Figure 1) to indicate whether the corresponding clause is satisfied. The data qubits (the qubits before Hadamard gates in Figure 1) represent the variables within a specific simplified clause. These variables are inputted into the C module to determine the feasibility of the corresponding clause. To ensure a feasible solution, all clauses (C modules) must be satisfied. Consequently, we combine two C modules sequentially using an AND circuit. More information about QFL circuit construction is provided with examples in section II-C and II-D.

**VQS**: Using the final labeling qubit *L* from the output of QFL, together with all possible solutions as input, VQS can find all feasible solutions of the simplified clauses, as detailed in the next three paragraphs. From the feasible solutions found by the VQS algorithm, we can extract the prime factors, denoted as *p* and *q*, for a given integer *N*.

In the context of the SDF method, a final labeling qubit, denoted as *L*, is utilized. This qubit assumes the state $|1\rangle$ for feasible solutions and $|0\rangle$ for infeasible ones. When combined with all possible solutions, this labeling qubit collectively represents an unstructured database. Each element in this database corresponds to a solution, accompanied by its feasibility label. A feasible solution is akin to a good element, while an infeasible solution corresponds to a bad element. This framework effectively links feasible and infeasible solutions to the concept of an unstructured database.

Grover's search algorithm (GSA) applies a negative phase to a good element to increase the probability of finding it. In contrast, VQS attaches a label state, $|1\rangle$, to the good elements. Both GSA and VQS employ this approach to amplify the likelihood of identifying the good elements within an unstructured database, as described in (Zhan 2022).

Considering the feasible and infeasible solutions are like good and bad elements in an unstructured database, as described above, VQS can amplify the probability of the feasible solutions, which is the last stage within the SDF. In other words, the VQS leverages the label state $|1\rangle$ on the final labeling qubit to amplify the probability of finding feasible solutions, just as it amplifies the probability of identifying good elements in an unstructured database.

**C. Circuit Construction for Example 1**

According to the last paragraph in Section II-A, both $n_p$ and $n_q$ are equal to 4 for *N*=143. It is noteworthy that since *N*=143 is odd, both *p* and *q* must be odd. The reason is that if either *p* or *q* were even, the product $N = p \times q$ would result in an even number. In order to ensure that *p* and *q* are odd, we set their least significant bit (LSB) (rightmost bit) to 1 as shown in the rightmost column of Table 2. By implementing binary multiplication, as explained in Section II-A, we can obtain equations (8)-(14) from Table 2:

$$p_1 + q_1 = 1 + 2z_{1,2} \quad (8)$$

$$p_2 + p_1 q_1 + q_2 + z_{1,2} = 1 + 2z_{2,3} + 4z_{2,4} \quad (9)$$

$$p_3 + p_2 q_1 + p_1 q_2 + q_3 + z_{2,3} = 1 + 2z_{3,4} + 4z_{3,5} \quad (10)$$

$$p_3 q_1 + p_2 q_2 + p_1 q_3 + z_{3,4} + z_{2,4} = 0 + 2z_{4,5} + 4z_{4,6} \quad (11)$$

$$p_3 q_2 + p_2 q_3 + z_{4,5} + z_{3,5} = 0 + 2z_{5,6} + 4z_{5,7} \quad (12)$$

$$p_3 q_3 + z_{5,6} + z_{4,6} = 0 + 2z_{6,7} \quad (13)$$

$$z_{6,7} + z_{5,7} = 1 \quad (14)$$

By applying the specific pre-processing rules outlined in Appendix A, equations (8) to (14) can be further simplified to be (15)-(18):

$$p_3 = 1, q_3 = 1 \quad (15)$$

$$p_1 + q_1 = 1 \quad (16)$$

$$p_2 + q_2 = 1 \quad (17)$$

$$p_1 q_2 + p_2 q_1 = 1 \quad (18)$$

Table 2: Binary Multiplication for Example 1.

|   | $2^7$ | $2^6$ | $2^5$ | $2^4$ | $2^3$ | $2^2$ | $2^1$ | $2^0$ |
|---|---|---|---|---|---|---|---|---|
| *p* |  |  |  |  | $p_3$ | $p_2$ | $p_1$ | 1 |
| *q* |  |  |  |  | $q_3$ | $q_2$ | $q_1$ | 1 |
|  |  |  |  |  | $p_3$ | $p_2$ | $p_1$ | 1 |
|  |  |  |  | $p_3 q_1$ | $p_2 q_1$ | $p_1 q_1$ | $q_1$ |  |
|  |  |  |  | $p_3 q_2$ | $p_2 q_2$ | $p_1 q_2$ | $q_2$ |  |
|  |  |  | $p_3 q_3$ | $p_2 q_3$ | $p_1 q_3$ | $q_3$ |  |  |
| *Carries* | $z_{6,7}$ | $z_{5,6}$ | $z_{4,5}$ | $z_{3,4}$ | $z_{2,3}$ | $z_{1,2}$ |  |  |
|  | $z_{5,7}$ | $z_{4,6}$ | $z_{3,5}$ | $z_{2,4}$ |  |  |  |  |
| *N*=143 | 1 | 0 | 0 | 0 | 1 | 1 | 1 | 1 |

We have designed the quantum circuit shown in Figure 2 to solve the factoring problem of 143, where $p_1, p_2, q_1, q_2$ are data qubits, used as inputs for modules $C_1, C_2$, and $C_3$, which correspond to (16)-(18), respectively. The qubits $L_1$ to $L_5$ represent the feasibility labeling qubits (FLQ), and $a_1$ denotes an ancillary qubit. The initial state of all qubits is $|0\rangle$.

In module $C_1$, the solution of (16) is evaluated, and depending on the output, labeling qubit $L_1$ is labeled with the state $|1\rangle$. Otherwise, $L_1$ is labeled with the state $|0\rangle$. Table 3 presents the truth table for the feasible solutions of module $C_1$. In module $C_1$, the first CNOT is used to realize the equation and store the result in $q_1$. Subsequently, a second CNOT gate is used to label the QFL qubit $L_1$. Finally, the third CNOT resets the input $q_1$, ensuring that its state remains unchanged before and after $C_1$ such that it can be reused in the subsequent C modules.

Table 3: QFL Truth Table for $C_1$ of Example 1.

| Input | | Output |
|---|---|---|
| $p_1$ | $q_1$ | $L_1$ |
| 0 | 0 | 0 |
| 0 | 1 | 1 |
| 1 | 0 | 1 |
| 1 | 1 | 0 |

Module $C_2$ is associated with (17). We replicate module $C_1$ to construct module $C_2$ because both modules require the same operations, but with different input variables. The FLQ qubit $L_2$ is in state $|1\rangle$ for feasible solutions, and in state $|0\rangle$ otherwise, following the execution of module $C_2$. To determine whether a solution satisfies both (16) and (17) or not,
we use an AND circuit that take the label qubits $L_1$ and $L_2$ as input. The label qubit $L_3$ stores the output of the AND circuit. When the label qubit $L_3$ is in the state $|1\rangle$, it indicates that both (16) and (17) are satisfied.

Module $C_3$ is associated with (18), which involves four variables. This equation is similar to (16) and (17) if we let $x = p_1 q_2$ and $y = p_2 q_1$. We apply binary multiplication to calculate $x$ and $y$. We employ an AND gate to achieve binary multiplication of two qubits. The first Toffoli of module $C_3$ calculates the product, $p_1 q_2$, and stores the result in ancillary $a_1$. The second Toffoli calculates $p_2 q_1$ and stores the result in $L_4$. Then, a CNOT is used to store the result of the whole module in $L_4$. Appendix B provides the truth table for module $C_3$. Since the current module is the final one, we do not reset the qubits for future use.

At last, we use an AND circuit to determine whether a solution satisfies (16), (17), and (18) by taking the label qubits $L_3$ and $L_4$ as input. The label qubit $L_5$ stores the output of the AND circuit. For any feasible solution that satisfies the three clauses (16)-(18), the quantum state associated with this solution should have $|1\rangle$ in qubit $L_5$. Then we apply VQS to extract all the feasible solutions from the possible solutions for $p_1, p_2, q_1, q_2$. We can determine

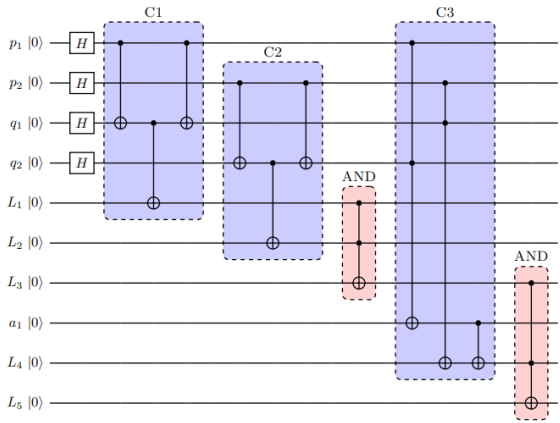

Figure 2: Quantum Feasibility Labeling Circuit for Example 1

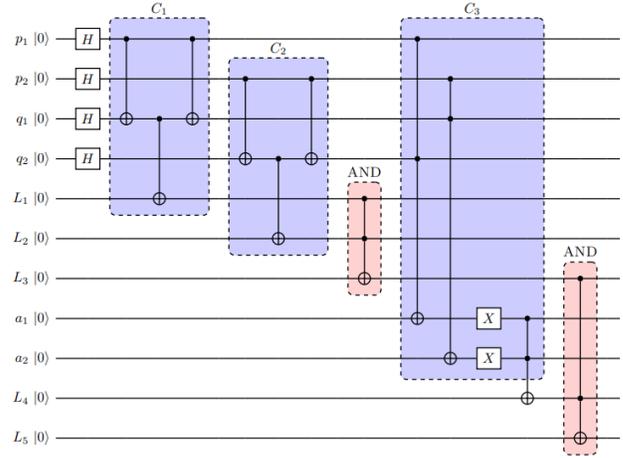

Figure 3: Quantum Feasibility Labeling Circuit for Example 2

the prime factors $p$ and $q$ using the values we get from VQS as described in Section III.

### D. Circuit Construction for Example 2

In this example, we consider $N=391$, a 9-bit number. According to the bit length estimation method described in Section II-A, we determine that both $p$ and $q$ are at most 5 bits in length. We can further simplify the equations we get from multiplication table (provided in Appendix C) by using the pre-processing rules described in Appendix A. This simplification process helps to reduce the complexity of the calculations involved. The simplified clauses derived from the multiplication table shown in Appendix C are as follows:

$$p_4 = 1, q_4 = 1, p_3 = 0, q_3 = 0 \quad (19)$$

$$p_1 + q_1 = 1 \quad (20)$$

$$p_2 + q_2 = 1 \quad (21)$$

$$p_1 q_2 + p_2 q_1 = 0 \quad (22)$$

The quantum circuit for factoring 391 is given in Figure 3. Equations (20)-(22) represent modules $C_1, C_2, C_3$ respectively. In Figure 3, $p_1, p_2, q_1, q_2$ are data qubits, used as inputs to modules. The feasibility labeling qubits are $L_1$ to $L_5$. The ancillary qubits are $a_1$ and $a_2$.

Since (20) is identical to (16), we employ the same $C_1$ circuit from Figure 2 to realize (20) in Figure 3. After evaluating (20) using module $C_1$, if the equation is satisfied, the QFL qubit $L_1$ will be in state $|1\rangle$.

We construct module $C_2$ in the same way as the $C_1$. This is because equations (20) and (21) are like each other. The label qubit $L_2$ stores the result of module $C_2$ in a similar manner to the previous module. After $C_2$, an AND circuit takes $L_1$ and $L_2$ as input and store the result in $L_3$. If $L_3$ is in state $|1\rangle$, it indicates that both (20) and (21) are satisfied.

Module $C_3$ in Figure 3 is to verify (22). There is no reset operation as this is the last module. The truth table for module $C_3$ is given in Table 4. The first two CCNOT gates in $C_3$ compute $p_1q_2$ and $p_2q_1$ from (22), and store outputs in ancillary qubits $a_1$ and $a_2$, respectively. The label qubit $L_4$ stores the result of module $C_3$ in a similar manner to the previous modules.

Table 4: QFL Truth Table for $C_2$ of Example 2.

| Input | | | | Output |
|---|---|---|---|---|
| $p_1$ | $p_2$ | $q_1$ | $q_2$ | $L_2$ |
| 0 | 0 | 0 | 0 | 1 |
| 0 | 0 | 0 | 1 | 1 |
| 0 | 0 | 1 | 0 | 1 |
| 0 | 0 | 1 | 1 | 1 |
| 0 | 1 | 0 | 0 | 1 |
| 0 | 1 | 0 | 1 | 1 |
| 0 | 1 | 1 | 0 | 0 |
| 0 | 1 | 1 | 1 | 0 |
| 1 | 0 | 0 | 0 | 1 |
| 1 | 0 | 0 | 1 | 0 |
| 1 | 0 | 1 | 0 | 1 |
| 1 | 0 | 1 | 1 | 0 |
| 1 | 1 | 0 | 0 | 1 |
| 1 | 1 | 0 | 1 | 0 |
| 1 | 1 | 1 | 0 | 0 |
| 1 | 1 | 1 | 1 | 0 |

We construct the final AND circuit to provide a logic AND to label qubits $L_3$ and $L_4$. The label qubit $L_5$ stores the output of final AND circuit. If the state of $L_5$ is $|1\rangle$, all 3 clauses (20)-(22) are satisfied. Then, we apply VQS to extract the feasible solutions for the clauses and determine the factors $p$ and $q$.

**E. Complexity Analysis**

To analyze the depth complexity of the SDF, we first calculate the complexity of the method without pre-processing. Let us consider integer $N$ with its prime factor $p$ and $q$. $n_N, n_p, n_q$ represent bit length of $N$, $p$ and $q$ respectively. The total number of equations is given by $n_c = n_p + n_q - 1$, resulting in a total of $n_c$ clauses. Among these, we focus on the clause with the highest number of terms, containing at most $n_a = n_k + 2\log_2(n_k) + 1$ terms, where $n_k$ is the maximum of $n_p$ and $n_q$. The terms refer to $p_iq_j$ and carries $z_{m,n}$. From the circuit construction, each term can have a circuit depth of at most 2. Hence, we require a maximum of $2n_a$ circuit depths to realize a single clause. Additionally, an extra $2n_a$ depths are needed for the reset operation, plus one for labeling qubit. Thus, each clause requires at most $4n_a + 1$ circuit depths. With a total of $n_c$ clauses, the maximum required depths become $n_c(4n_a + 1)$. Furthermore, $2n_c - 1$ circuit depths are needed for the AND operation, along with one for the Hadamard operation. In total, the maximum circuit depth is $n_c(4n_a + 1) + 2n_c$. The **complexity** of the circuit depth can be written as $\mathcal{O}(8n_k^2)$ or $\mathcal{O}(8(\log_2 N)^2)$. However, this paper uses pre-processing rules, which are anticipated to yield a substantial reduction in circuit depth. For instance, in example 1, the calculated circuit depth without pre-processing rules is 217, whereas implementing pre-processing rules brings it down to only 12. Again, in example 2, the circuit depth is 315, but applying pre-processing rules reduces it to just 13. This implies that the proposed method offers the advantage of a reduced circuit depth.

Here we consider the complexity in terms of qubit utilization. The number of data qubits required is $n_p + n_q$. The bilinear terms in equation (4) require a maximum of $n_p \times n_q$ ancillary qubits. The upper bound for the number of carry qubits is $n_a \times log_2(n_k)$. There are $2n_c - 1$ label qubits. In total, all these qubits add up to $n_p + n_q + n_pn_q + n_a \log_2(n_k) + 2n_c - 1$. The **complexity** of the number of qubits can be written as $\mathcal{O}(n_pn_q)$.

## III. Results

We implemented the quantum algorithm using PennyLane (Bergholm et al. 2018). For the case of *N*=143, the circuit depicted in Figure 2 was executed, and VQS was applied to determine the feasible solution using the label qubit $L_5$ and the corresponding states as input. Following execution of the circuit in Figure 2 and VQS, we got the feasible solutions (indicated by $L_5 = 1$) which are tabulated in Table 5.

Table 5: Results of Example 1 Obtained from Execution of SDF Algorithm

| $p_1$ | $p_2$ | $q_1$ | $q_2$ | $L_5$ |
|---|---|---|---|---|
| 1 | 0 | 0 | 1 | 1 |
| 0 | 1 | 1 | 0 | 1 |

Based on Table 2, we express $p$ and $q$ as:

$$p = (p_3p_2p_11)_2$$

$$q = (q_3q_2q_11)_2$$

From Table 5, we have two feasible solutions of $p$ and $q$ when $L_5$ is 1. Using both solutions, we get 11 and 13 as the factors of 143. Taking the first solution of $p$ and $q$ from Table 5 (the 2nd row), we obtain:

$$p = 1011_2 = 11$$

$$q = 1101_2 = 13$$

Note that, the values of $p_3$, $q_3$ are taken from (15).

For *N*=391, after executing the circuit in Figure 3 and then the VQS, we obtain the feasible solutions, following the same method as for *N*=143. The feasible solutions are listed in Table 6.

According to Appendix C, we express $p$ and $q$ as:

$$p = (p_4 p_3 p_2 p_1 1)_2$$

$$q = (q_4 q_3 q_2 q_1 1)_2$$

Table 6: Results of Example 2 Obtained from Execution of SDF Algorithm

| $p_1$ | $p_2$ | $q_1$ | $q_2$ | $L_5$ |
|---|---|---|---|---|
| 0 | 0 | 1 | 1 | 1 |
| 1 | 1 | 0 | 0 | 1 |

Using the first feasible solutions in Table 6 (the 2$^{nd}$ row), we can obtain the factors of 391 as below:

$$p = 10111_2 = 23$$

$$q = 10001_2 = 17$$

Taking the second solution from Table 6, we obtain:

$$p = 10001_2 = 17$$

$$q = 10111_2 = 23$$

where the values of $p_4$, $q_4$, $p_3$ and $q_3$ are taken from (19). Both feasible solutions from Table 6 yield 17 and 23 as the factors of 391.

## IV. Conclusion

Integer factoring has been an important area of research for the last few decades because of its application in public key cryptosystems. In this paper, we propose a new method, SDF, for integer factoring. Using the labels obtained by the QFL and all the corresponding possible solutions as input, the VQS can find all feasible solutions for the clauses of the factoring problem. Based on these feasible solutions, we get factors of the given integer. The SDF algorithm utilizes shallow-depth quantum circuits for efficient factorization, with the complexity of the circuit depth being $\mathcal{O}(8(\log_2 N)^2)$. The complexity of the number of qubits is $\mathcal{O}(n_p n_q)$. While the method's progress in reducing the number of qubits remains limited, it notably offers a significant advantage in terms of circuit depth. Further advancements in reducing the number of qubits represent a promising direction for future research exploration. Through minimizing the number of gates in the circuit, the algorithm enhances feasibility and reduces vulnerability to errors. Our proposed SDF method shows promise for factoring large integers, holding implications for the field of cryptography.

## References

Anschuetz, E; Olson, J; Aspuru-Guzik, A; Cao, Y. 2019. "Variational Quantum Factoring." In *Quantum Technology and Optimization Problems: First International Workshop, QTOP 2019, Munich, Germany, March 18, 2019, Proceedings 1*, 74–85. Springer. https://doi.org/10.1007/978-3-030-14082-3_7

Aoki, K; Franke, J; Kleinjung, T; Lenstra, AK; Osvik, DA. 2007. "A Kilobit Special Number Field Sieve Factorization." In *Advances in Cryptology–ASIACRYPT 2007: 13th International Conference on the Theory and Application of Cryptology and Information Security, Kuching, Malaysia, December 2-6, 2007. Proceedings 13*, 1–12. Springer. https://doi.org/10.1007/978-3-540-76900-2_1

Bergholm, V; Izaac, J; Schuld, M; Gogolin, C; Ahmed, S. et al. 2018. "PennyLane: Automatic Differentiation of Hybrid Quantum-Classical Computations." *ArXiv* abs/1811.04968.

Biamonte, J; Wittek, P; Pancotti, N; Rebentrost, P; Wiebe, N; Lloyd, S. 2016. "Quantum Machine Learning." *Nature* 549: 195–202. https://doi.org/10.1038/nature23474

Burges, C J C. 2002. "Factoring as Optimization." *Microsoft Research MSR-TR-200*.

Cai, W; Ma, Y; Wang, W; Zou, CL; Sun, L. 2021. "Bosonic Quantum Error Correction Codes in Superconducting Quantum Circuits." *Fundamental Research* 1 (1): 50–67. https://doi.org/10.1016/j.fmre.2020.12.006.

Cao, Y; Romero, J; Olson, JP; Degroote, M; Johnson, PD; Kieferová, M; Kivlichan, D et al. 2019. "Quantum Chemistry in the Age of Quantum Computing." *Chemical Reviews* 119 (19): 10856–915. https://doi.org/10.1021/acs.chemrev.8b00803.

Dash, A; Sarmah, D; Behera, BK; Panigrahi, PK. 2018. "Exact Search Algorithm to Factorize Large Biprimes and a Triprime on IBM Quantum Computer," May. http://arxiv.org/abs/1805.10478.

Euclid. 2002. *Euclid's Elements*. Edited by Densmore D. New Mexico: Green Lion Press.

Farhi, E; Goldstone, J; Gutmann, S; Sipser, M. 2000. "Quantum Computation by Adiabatic Evolution." arXiv:quant-ph/0001106

Gamel, O, James, DFV. 2013. "Simplified Factoring Algorithms for Validating Small-Scale Quantum Information Processing Technologies," May. http://arxiv.org/abs/1310.6446.

Geller, MR, Zhou, Z. 2013. "Factoring 51 and 85 with 8 Qubits." *Scientific Reports* 3. https://doi.org/10.1038/srep03023

Georgescu, IM; Ashhab, S; Nori, F. 2014. "Quantum Simulation." *Reviews of Modern Physics* 86 (1): 153–85. https://doi.org/10.1103/RevModPhys.86.153.

Grover, LK. 1996. "A Fast Quantum Mechanical Algorithm for Database Search." In *Symposium on the Theory of Computing*. https://doi.org/10.1145/237814.237866

Hardy, GH, Wright EM. 1975. "An Introduction to the Theory of Numbers." London: Oxford University Press.

Lenstra, AK, Lenstra HW. 1993. *The Development of the Number Field Sieve*. Berlin: Springer Science & Business Media.

Long, CT. 1972. "Elementary Introduction to Number Theory." *Lexington: DC Heath and Company, LCCN* 77171950: 7.

McArdle, S; Jones, T; Endo, S; Li, Y; Benjamin, SC; Yuan, X. 2018. "Variational Quantum Simulation of Imaginary Time Evolution with Applications in Chemistry and Beyond." In npj Quantum Information 5, 75. https://doi.org/10.1038/s41534-019-0187-2

Mollin, RA. 2002. "A Brief History of Factoring and Primality Testing BC (before Computers)." *Mathematics Magazine* 75 (1): 18–29. https://doi.org/10.1080/0025570X.2002.11953094

Ortiz, G; Gubernatis, JE; Knill, E; Laflamme, R. 2001. "Quantum Algorithms for Fermionic Simulations." *Physical Review A* 64 (2): 22319. https://doi.org/10.1103/PhysRevA.64.022319.

Peng, X; Liao, Z; Xu, N; Qin, G; Zhou, X; Suter, D; Du, J. 2008. "Quantum Adiabatic Algorithm for Factorization and Its Experimental Implementation." *Physical Review Letters* 101 (22): 220405. https://doi.org/10.1103/PhysRevLett.101.220405

Peruzzo, A; McClean, J; Shadbolt, P; Yung, MH; Zhou, XQ; Love, PJ; Aspuru-Guzik, A; O'Brien, JL. 2014. "A Variational Eigenvalue Solver on a Photonic Quantum Processor." *Nature Communications* 5 (1): 4213. https://doi.org/10.1038/ncomms5213.

Pettofrezzo, AJ, Byrkit, DR. 1970. *Elements of Number Theory*. New Jersey: Prentice-Hall.

Pollard, JM. 1975. "A Monte Carlo Method for Factorization." *BIT Numerical Mathematics* 15 (3): 331–34. https://doi.org/10.1007/BF01933667.

Pomerance, C. 1982. "Analysis and Comparison of Some Integer Factoring Algorithms." *Computational Methods in Number Theory*, 89–139. Amsterdam: Mathematich Centrum.

Reed, MD; DiCarlo, L; Nigg, SE; Sun, L; Frunzio, L; Girvin, SM; Schoelkopf, RJ. 2012. "Realization of Three-Qubit Quantum Error Correction with Superconducting Circuits." *Nature* 482 (7385): 382–85. https://doi.org/10.1038/nature10786.

Shor, PW. 1994. "Algorithms for Quantum Computation: Discrete Logarithms and Factoring." In *Proceedings 35th Annual Symposium on Foundations of Computer Science*, 124–34. https://doi.org/10.1109/SFCS.1994.365700.

Skosana, U, Tame, M. 2021. "Demonstration of Shor's Factoring Algorithm for N = 21 on IBM Quantum Processors." *Scientific Reports* 11 (1). https://doi.org/10.1038/s41598-021-95973-w.

Stamatopoulos, N; Egger, D; Sun, Y; Zoufal, C; Iten, R; Shen, N; Woerner, S. 2020. "Option Pricing Using Quantum Computers." *Quantum* 4 (July): 291. https://doi.org/10.22331/q-2020-07-06-291.

Wang, B; Hu, F; Yao, H; Wang, C. 2020. "Prime Factorization Algorithm Based on Parameter Optimization of Ising Model." *Scientific Reports* 10 (1). https://doi.org/10.1038/s41598-020-62802-5.

Xu, N; Zhu, J; Lu, D; Zhou, X; Peng, X; Du, J. 2011. "Quantum Factorization of 143 on a Dipolar-Coupling NMR System." In Phyical Review Letter. 109, 269902 (2012). https://doi.org/10.1103/PhysRevLett.108.130501

Yan, B; Tan, Z; Wei, S; Jiang, H; Wang, W; Wang, H et al. 2022. "Factoring Integers with Sublinear Resources on a Superconducting Quantum Processor." *ArXiv Preprint ArXiv:2212.12372*.

Yost, DRW; Schwartz, ME; Mallek, J; Rosenberg, D; Stull, C; Yoder, JL; Calusine, G et al. 2020. "Solid-State Qubits Integrated with Superconducting through-Silicon Vias." *Npj Quantum Information* 6 (1): 59. https://doi.org/10.1038/s41534-020-00289-8.

Zhan, J. 2022. "Variational Quantum Search with Shallow Depth for Unstructured Database Search." *ArXiv Preprint ArXiv:2212.09505*.

Zhan, J. 2023a. "Near-Perfect Reachability of Variational Quantum Search with Depth-1 Ansatz." *ArXiv Preprint ArXiv:2301.13224*.

Zhan, J. 2023b. "Quantum Feasibility Labeling for NP-Complete Vertex Coloring Problem." *ArXiv Preprint ArXiv:2301.01589*.

## ACKNOWLEDGMENTS

This research was partially supported by the NSF ERI program, under award number 2138702. This work used the Delta system at the National Center for Supercomputing Applications through allocation CIS220136 from the Advanced Cyberinfrastructure Coordination Ecosystem: Services & Support (ACCESS) program, which is supported by National Science Foundation grants #2138259, #2138286, #2138307, #2137603, and #2138296. We acknowledge the use of IBM Quantum services for this work. The views expressed are those of the authors, and do not reflect the official policy or position of IBM or the IBM Quantum team.

## Appendix

### A. Preprocessing Rules

Pre-processing rules are applied to simplify the clauses from the multiplication table. The carries calculated by (4), depend on the values of the terms in the first part ($\sum_{j=0}^{i} q_j p_{i-j}$), which can be either 1 or 0. In the case of 2 and 3 terms, we normally need 1 carry bit along with 1 LSB. This is because the binary addition of 2 and 3 terms results at most in the binary representations of 10 and 11, respectively, which require 2 bits. We do not need any carry bit for only 1 term as binary of 1 is 1 which requires only 1 bit. That is why the two rightmost columns of Table 1 do not have carries. For 4 to 7 terms, we need 2 carry bits along with 1 LSB (denoted as $N_i$ in Table 1) as binary representation of 4 to 7 is 100 to 111 (3 bits). Therefore, for example 1, we assign one carry ($z_{1,2}$) for $i = 2$ and two carries (i.e., $z_{2,3}$ and $z_{2,4}$) for $i > 2$ and the remaining terms of $z_{j,i}$ terms of the $\sum_{j=0}^{i} z_{j,i}$ expression from (4) are set to zero. We derive (8) as the first equation from the second rightmost column of Table 2. In equation (8), the term 1 on the right-hand side represents the LSB of the sum ($p_1 + q_1$). The most significant bit (MSB), which is the carry bit, is denoted as $z_{1,2}$. When adding two binary digits, the possible outcomes for their sum are: 00 (both digits are 0), 01 (one digit is 1 and the other is 0), or 10 (both digits are 1, resulting in a carry bit of 1). We can know that when the LSB of sum is 1, the MSB is 0. Since the LSB of the sum of $p_1$ and $q_1$ in (8) is 1, the MSB $z_{1,2}$ (carry bit) must be 0. This simplifies (8) to (16).

Indeed, if the sum ($p_1 + q_1$) is equal to 1 in binary representation, it implies that one of the bits must be 0, while the other is 1. Therefore, it follows that the product $p_1*q_1$ must be 0, as the multiplication of 0 with any other bit yields 0. We write the rule for binary as:

$$x + y = 1, \text{ then } xy = 0 \qquad (23)$$

We derive (9) as the second equation for example 1 from the third rightmost column of Table 2. Here, the LSB is 1 (from the right-hand side of (9)), $z_{2,3}$ is the second LSB, and $z_{2,4}$ is the MSB. As per previous deduction, $p_1 q_1 = 0$ and $z_{1,2} = 0$, we can convert (9) to:

$$p_2 + 0 + q_2 + 0 = 1 + 2z_{2,3} + 4z_{2,4} \qquad (24)$$

If we add 4 binary digits based on the left-hand side of (24), we can get one of 000, 001, 010 as a result. Matching with these possible results, we can infer that when the LSB of the sum is 1, the second LSB and MSB both are 0. Since the LSB of the sum of 4 terms of left-hand side of (24) is 1, the second LSB and the MSB both are 0. Hence, both the second LSB, $z_{2,3}$, and the MSB, $z_{2,4}$, are set to 0 and we obtain (17) as a simplified version of (24).

Similarly, we use following binary pre-processing rules to get the simplified clauses:

$$xy = 1, \rightarrow x = y = 1 \qquad (25)$$

$$\sum_i x_i = 0, \rightarrow x_i = 0 \qquad (26)$$

$$x + y = 2z, \rightarrow x = y = z \qquad (27)$$

$$x + 2y - 2z = 0, \rightarrow x = 0, y = z \qquad (28)$$

$$x - 2z + 1 = 0, \rightarrow x = 1, z = 1 \qquad (29)$$

The pre-processing rules mentioned above serve as examples to simplify the factoring problem into an optimization problem. One can produce their own pre-processing rules as needed. By applying these rules, the clauses become easier to solve.

**B. QFL Truth Table for Equation (18)**

Table 7 is the truth table for module $C_3$ of example 1 (Factoring 143).

Table 7: QFL Truth Table for $C_3$ of Example 1.

| Input | | | | Output |
|---|---|---|---|---|
| $p_1$ | $p_2$ | $q_1$ | $q_2$ | $L_2$ |
| 0 | 0 | 0 | 0 | 0 |
| 0 | 0 | 0 | 1 | 0 |
| 0 | 0 | 1 | 0 | 0 |
| 0 | 0 | 1 | 1 | 1 |
| 0 | 1 | 0 | 0 | 0 |
| 0 | 1 | 0 | 1 | 0 |
| 0 | 1 | 1 | 0 | 0 |
| 0 | 1 | 1 | 1 | 1 |
| 1 | 0 | 0 | 0 | 0 |
| 1 | 0 | 0 | 1 | 0 |
| 1 | 0 | 1 | 0 | 0 |
| 1 | 0 | 1 | 1 | 1 |
| 1 | 1 | 0 | 0 | 1 |
| 1 | 1 | 0 | 1 | 1 |
| 1 | 1 | 1 | 0 | 1 |
| 1 | 1 | 1 | 1 | 0 |

**C. Binary Multiplication Table for Example 2**

Table 8 presents the binary multiplication table for the integer 391. $C$ stands for carries in Table 8. Similar to Table 2, both $p$ and $q$ must be odd since $N=391$ is odd. Thus, we set $p_0$ and $q_0$ as 1.

Table 8: Binary Multiplication for Example 2 (N=391).

| | $2^9$ | $2^8$ | $2^7$ | $2^6$ | $2^5$ | $2^4$ | $2^3$ | $2^2$ | $2^1$ | $2^0$ |
|---|---|---|---|---|---|---|---|---|---|---|
| $p$ | | | | | | $p_4$ | $p_3$ | $p_2$ | $p_1$ | 1 |
| $q$ | | | | | | $q_4$ | $q_3$ | $q_2$ | $q_1$ | 1 |
| | | | | | | $p_4$ | $p_3$ | $p_2$ | $p_1$ | 1 |
| | | | | | $p_4 q_1$ | $p_3 q_1$ | $p_2 q_1$ | $p_1 q_1$ | $q_1$ | |
| | | | | $p_4 q_2$ | $p_3 q_2$ | $p_2 q_2$ | $p_1 q_2$ | $q_2$ | | |
| | | | $p_4 q_3$ | $p_3 q_3$ | $p_2 q_3$ | $p_1 q_3$ | $q_3$ | | | |
| | | $p_4 q_4$ | $p_3 q_4$ | $p_2 q_4$ | $p_1 q_4$ | $q_4$ | | | | |
| $C$ | $z_{8,9}$ | $z_{7,8}$ | $z_{6,7}$ | $z_{5,6}$ | $z_{4,5}$ | $z_{3,4}$ | $z_{2,3}$ | $z_{1,2}$ | | |
| | $z_{7,9}$ | $z_{6,8}$ | $z_{5,7}$ | $z_{4,6}$ | $z_{3,5}$ | $z_{2,4}$ | | | | |
| N=391 | 0 | 1 | 1 | 0 | 0 | 0 | 0 | 1 | 1 | 1 |